# Configurable Vibrational Coupling in Laser-Induced Microsecond Oscillations of Multi-Microbubble System

Xuanwei Zhang[1], Ryu Matsuo[1], Jun Nishida[2,3], Kyoko Namura[1], Motofumi Suzuki[1]

[1]Department of Micro Engineering, Kyoto University, Kyoto Daigaku-Katsura, Nishikyo-ku, Kyoto 615-8540, Japan

[2]Institute for Molecular Science, National Institutes of Natural Sciences, Okazaki, Aichi, 444-8585, Japan

[3]The Graduate Institute for Advanced Studies, SOKENDAI, Hayama, Kanagawa, 240-0193, Japan

# Abstract

We study the coupled vibrational dynamics of sub-MHz self-oscillating bubbles at separations of 14 to 92 μm. Two vapor-rich microbubbles are generated via photothermal heating; their interactions are captured in real space and time via a high-speed camera. By controlling the distance between bubbles with micrometer precision, we induce in- and anti-phase hybridized vibrations, with mode frequencies varying from 0.5 to 0.8 MHz. The observed coupled modes are reproduced by the extended Rayleigh-Plesset equation that takes pressure interactions between the bubbles into account. Our findings facilitate the optimization of bubble array positioning for spatio-temporal microfluidic control.

# Introduction

Micro- and nano-bubbles dynamically change their volumes through phase changes and gas compression, which in turn drive fluids and emit acoustic waves. These properties are valuable for medical applications [1-8], microfluidics [9-15], and chemical processes [16,17]. The dynamics of single and isolated bubbles have been extensively studied and are well understood [18-20], along with the resulting acoustic emission [21] and flow generation [21-25]. Extending such single-bubble systems to multi-bubble systems would naturally facilitate stronger acoustic waves and spatially patterned flows. Ordered bubble arrays have been shown to generate complex flows [26,27], drive objects [28,29], and control ultrasound waves [30]. However, the underlying microscopic interactions among bubbles and the principles governing the complex flows generated around them remain elusive.

The vibrational interactions of bubbles have been extensively studied in terms of resonant frequency variations due to acoustic coupling [31-34]. Previous research demonstrated that analytical frequency solutions agree well with experimental results for large air-rich bubbles (approximately 1 mm in radius and < 10 kHz in frequency), which vibrate at small amplitudes [35-37]. However, the size and resonant frequency of these bubbles differ significantly from those relevant to microfluidic applications. Meanwhile, recent advances in microheaters have facilitated the creation of vapor-rich microbubbles that oscillate at large amplitudes [38-41]. Upon heating water with 10-μm heaters, vapor-rich bubbles of comparable size are generated, self-oscillating at sub-MHz frequencies and driving flows at speeds up to 1 m/s [42-44]. By arranging these bubbles on a micrometer scale and enabling them to interact strongly with each other, spatiotemporally complex and strong flows are expected to be created in the microfluidic channels. In recent progress, Nguyen et al. observed synchronized vibrations between two vapor-rich bubbles at distances of 40–50 μm, emphasizing the relationship between bubble distance and interaction strength [45]. However, the use of micro-structured heaters with fixed configurations and non-degassed water complicates the control over oscillation and leaves the exploration of coupling regimes limited.

Here, we demonstrate configurable vibrational coupling between two self-oscillating and vapor-rich microbubbles in degassed water. Using a spatial light modulator (SLM) for multi-point laser irradiation, the bubble positions are controlled on a photothermal conversion thin film, and their separation distances are varied systematically. Degassing the water ensures stable oscillations and prevents the cessation of bubble oscillations. We capture vibrational interactions using a 5 Mfps ultra-high-speed camera, which directly reveals large-amplitude vibrations which are hybridized into synchronous in- and anti-phase vibration modes at sufficiently short distances. The coupling strength between the bubbles is readily regulated by adjusting the bubble-bubble distances, resulting in the tuning of resonance frequency between 0.5 and 0.8 MHz. The vibration frequencies and their distance dependence are quantitatively reproduced using the extended Rayleigh-Plesset equation, assuming the perturbative limit despite the large amplitude oscillations.

## Method

The amorphous $FeSi_2$ thin film used for photothermal heating was formed on a glass substrate via radio frequency magnetron sputtering at room temperature. The fabrication method is described in our previous study [42]. The thickness of the $FeSi_2$ thin film was tuned to 50 nm by optimizing the deposition time. The optical absorptance was 30.7 % at a wavelength of 830 nm (see Supplemental Material, S1 [46]), at which point the photothermal conversion experiments were conducted.

The prepared $FeSi_2$ thin film was cut into small piece and attached to the internal wall of a glass cell (10 mm × 10 mm × 58 mm, F15-G-10, GL Science) using a pair of magnets. The cell was filled with degassed water, prepared via the sonication of ultrapure water (18.2 MΩ cm form Millipore-Direct Q UV3, Merck) under vacuum for at least 20 min. The cell was then sealed to prevent non-condensable gases from diffusing back into the water. Bubbles were generated within 1 h of cell preparation to conduct the experiment under conditions of low non-condensable gas concentrations in water.

The prepared fluidic cell was positioned within the optical setup (see Supplemental Material, S2 [46]). To facilitate the formation of microbubbles and induce their oscillations, a continuous-wave laser output (FPL830S, Thorlabs) with a wavelength of 830 nm was directed onto the FeSi$_2$ thin film. Prior to the irradiation of the thin film, the laser beam was modulated using an SLM (X10468-02, Hamamatsu Photonics), which facilitated the simultaneous irradiation of multiple laser spots and control of the distance between these laser spots [42]. As a result, two laser spots were observed on the FeSi$_2$ thin film, and a microbubble was generated on each spot. An ultra-high-speed camera (HPV-X2, Shimadzu) operating at a frame rate of 5 Mfps was utilized to capture the oscillations of the microbubbles from a direction almost parallel to the side of the substrate surface. The camera was synchronized with an ultra-high-speed laser diode (640 nm, Cavilux Smart UHS, Cavitar Inc.) with a 20 ns pulse width, which was utilized to illuminate the bubbles. Each shot was recorded in 255 frames and lasted 51 μs. After each recording, laser irradiation position was changed to mitigate potential damage to the film structure.

# Results and Discussion

Before investigating the vibrational interaction between two bubbles, we examined the relationship between the size and vibrational frequency of a single bubble. First, a single laser spot was generated on the FeSi$_2$ thin film. This resulted in the formation of a vapor-rich bubble in the degassed water owing to the photothermal effect of the FeSi$_2$ film [Fig. 1(a)]. This bubble underwent self-oscillations of the sub-MHz order under CW laser irradiation [40]. The oscillatory behavior of the bubble was captured via an ultra-high-speed camera (see Supplemental Material, S3 and Movie S1 [46]). The temporal evolution of the bubble radius $R(t)$ in the direction parallel to the substrate surface is shown in the inset of Fig. 1(b). A stable oscillation of the radius of 2.0–5.5 μm was obtained. A Fourier transform of the time-domain profile yielded a peak at the frequency of 0.55 MHz [Fig. 1(b)]. Higher harmonic peaks (1.1, 1.7, and 2.2 MHz) were also observed owing to the non-harmonic profile of the temporal oscillation. In the following discussion, we focus on the fundamental oscillation at 0.55 MHz. The derived oscillation frequency agrees well with the intensity variation of the scattered light from the bubble measured in a previous study (see Supplemental Material, S4 [46]).

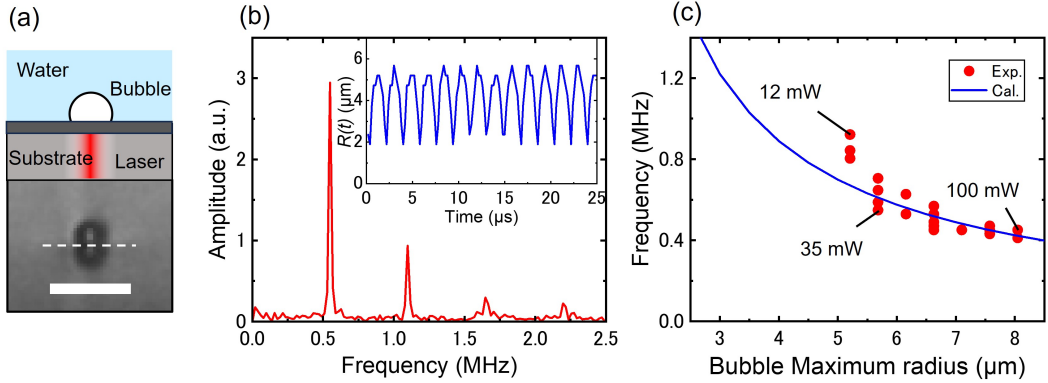

FIG.1. Observation of the single microbubble oscillation. (a) Typical microscopic images of the microbubble oscillation. Scale bar, 30 μm (b) Temporal evolution of the bubble radius $R(t)$ (inset, blue line) and Fourier transform results (red line). (c) Red circles indicate the experimentally obtained relationship between bubble maximum radius and oscillation frequency, and the blue line shows the calculated resonance frequency of the bubbles by Rayleigh-Plesset model.

We then varied the incident laser power from 12 to 100 mW and found an inverse correlation between the maximum bubble radius and oscillation frequency, which is consistent with a previous report [40] [Fig. 1(c)] (for further experimental details, see Supplemental Material, S5 [46]). The linear resonance frequency calculated at different $R_0$ based on the Rayleigh-Plesset model [18,19,47] is plotted as a blue line in Fig. 1(c) (see the Supplemental Material, S6 [46]). Regardless of the relatively large oscillation amplitudes in our microbubbles, the frequencies calculated based on the adiabatic and small-amplitude oscillation approximations reproduce our experimental results very well. This justifies the use of the same scheme and approximations to analyze the coupled oscillations between two neighboring microbubbles.

We then introduced two symmetrical laser focal spots on the photothermal conversion film using an SLM (each with a power of 35 mW) to simultaneously induce two vapor-rich microbubbles at each laser spot. By precisely regulating the distance between the laser spots ($d$), which corresponds to the bubble separation, we observed a clear coupling phenomenon between the microbubble oscillators. The real-space dynamics of the two bubbles recorded by an ultra-high-speed camera are illustrated in Fig. 2 (see Supplemental Material, Movie S2–S4 [46]). Fig. 2(a) shows the time-dependent variation in the radii of the two bubbles with a bubble distance of 67.1 μm. The red and blue lines represent the motions of the bubbles on the left and right sides, respectively. The inset shows a snapshot of the bubbles during vibration. The Fourier transform of the radius temporal profile [Fig. 2(b)] yielded a peak at 0.50 MHz, which is close to the value of 0.55 MHz (dotted line) observed in the case of a single isolated microbubble [Fig. 1(b)]. Interestingly, the two bubbles oscillated in sync, similar to observations by Nguyen et al. in their weakly interacting bubbles [45]. Achieving strong coupling between the bubbles requires further reduction in the distance, which has so far remained elusive due to the lack of facile control over inter-bubble distance on the order of 1 μm and the influence of dissolved gases. Here, with the superb control of laser spots offered by the SLM, along with the degassed water to mitigate the bubble growth and merging, we successfully reduced the distance to 27.6 μm and induced the bubbles' oscillations in a strong coupling regime [Fig. 2(c)]. From the individual radius variations of the two bubbles over time, we observed the occurrence of beats. This indicated the superposition of two coherent vibrations at different frequencies. In the Fourier transformed spectrum [Fig. 2(d)], two frequency peaks were observed at 0.50 and 0.62 MHz. When the two bubbles were in further proximity ($d$ = 13.9 μm), as shown in Fig. 2(e), they exhibited anti-phase vibrations. As shown in Fig. 2(f), the frequencies of the two bubbles increased to 0.81 MHz, which was a significant increase in the resonance frequency compared to isolated bubbles. Thus bubble-bubble interactions and the associated hybridization fundamentally alter the nature of the vibrations.

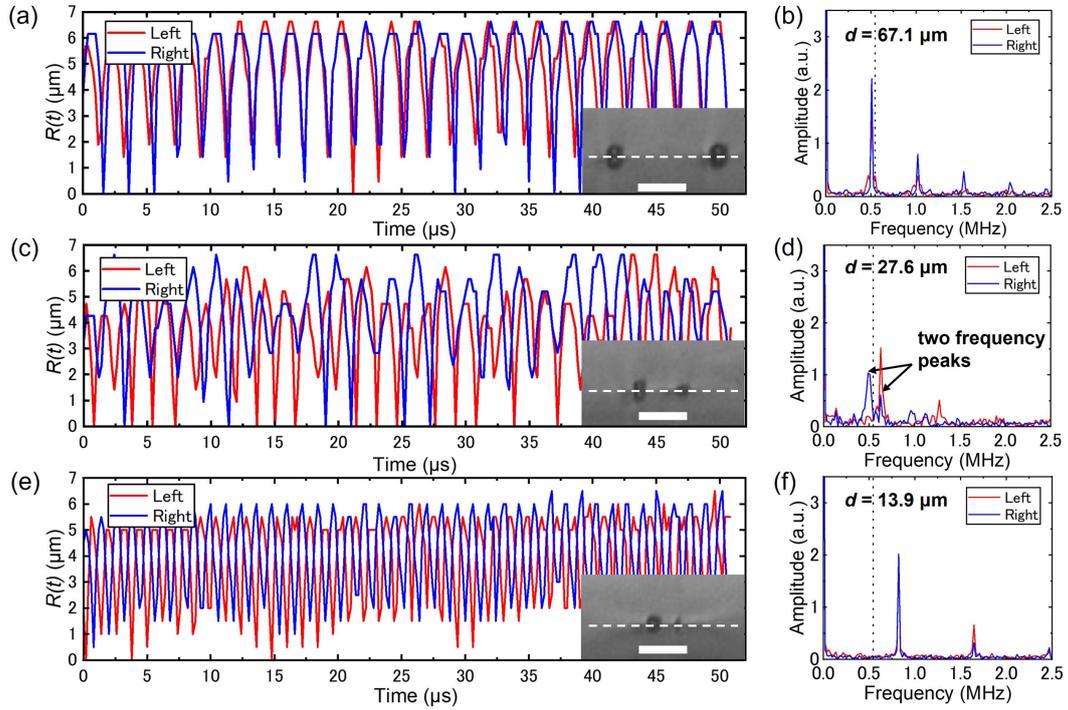

FIG.2. Observation of the coupling microbubble oscillation. Variation in the radius of the two bubbles across all 255 frames and the results of the Fourier transform of the radii variations of the two bubbles over time when (a, b) $d$ = 67.1 μm, (c, d) $d$ = 27.6 μm and (e, f) $d$ = 13.9 μm. Scale bars in a, c and e are 30 μm.

Through systematical adjustments of the distances of the laser point using the SLM, we analyzed the oscillation spectra of the left and right bubbles individually at different distances between the bubbles (see Supplemental Material, S7 [46]). At the distances smaller than 13.9 μm, the two bubbles tended to merge into one bubble. Thus, we only discuss the range of d ≥ 13.9 μm in this study. To analyze the vibrational coupling of interacting microbubble oscillators, the phase difference must be appropriately characterized. First, Fourier transforms were performed on the time-varying radii of the right and left bubbles, respectively. We then calculated the amplitude and phase of each bubble's oscillation at each frequency. Next, we computed the phase difference between the two bubbles at each bubble distance and each oscillation frequency and represented it in a color map, as shown in Fig. 3. The phase difference was wrapped within the range of 0–π. We masked the region with the lower-amplitude spectrum. The red domains indicate anti-phase vibrations, whereas the blue regions indicate in-phase vibrations.

As depicted in Fig. 3, when $d$ is between 53 and 92 μm, the interaction and mode hybridization is already present between the bubbles; however, their frequencies are very similar to each other and to that of an isolated bubble. At around $d = 60$ μm the presence of in-phase synchronization became pronounced. At $d = 53$–$23$ μm, the two bubbles interact strongly to form distinct hybridized modes, with an anti-phase mode at the higher frequency and an in-phase mode at the lower frequency. When the distance between the bubbles was further decreased to $d = 23$–$14$ μm, the vibration was dominated by the anti-phase mode with negligible amplitude from the in-phase mode.

To understand this coupling behavior quantitatively, we theoretically calculated the resonance frequencies of the bubbles in the in- and anti-phase modes using the extended Rayleigh-Plesset model, which takes pressure interactions into account. By considering the interaction between the bubbles based on the modification of the pressure field induced by each individual bubble, the radial dynamics $R_i$ of each bubble is governed by a multi-bubble Rayleigh-Plesset model [48,49]:

$$R_i \ddot{R}_i + \frac{3}{2}\dot{R}_i = \frac{1}{\rho}\left(p_{bi0}\left(\frac{R_{i0}}{R_i}\right)^{3\varkappa} - p_\infty - \frac{2\sigma}{R_i} - \frac{4\mu}{R_i}\dot{R}_i\right) - \sum_{j\neq i}\frac{R_j^2 \ddot{R}_j + 2R_j\dot{R}_j^2}{d_{ij}} \quad (1)$$

$$p_{bi0} = \frac{2\sigma}{R_{i0}} + p_\infty \quad (2)$$

where $R_{i,j}$ is the radius of bubbles $i$ and $j$, $R_{i0}$ is the equilibrium radius of bubble $i$, $p_{bi0}$ is the bubble pressure of bubble $i$, $d_{ij}$ is the distance between the bubble $i$ and $j$, $\varkappa = {^{c_p}}/{c_v}$ is the polytropic exponent, $\sigma$ is the surface tension, $\mu$ is the dynamic viscosity, $\rho$ is the liquid density, and $p_\infty$ is the ambient pressure.

By assuming perturbative limit and following the method outlined in supplementary information S6 [46], we adopt $R_1 = R_2 = R_0(1 + \varepsilon)$ for the in-phase mode and $R_1 = R_0(1 + \varepsilon)$, $R_2 = R_0(1 - \varepsilon)$ for the anti-phase mode. Substituting these values into Eq. 1 while retaining $\varepsilon$ up to the first order, we obtain:

$$\left(R_0^2 \pm \frac{R_0^3}{d_{12}}\right)\ddot{\varepsilon} + \frac{4\mu}{\rho}\dot{\varepsilon} + \frac{1}{\rho}\left(3\varkappa p_\infty + \frac{2\sigma}{R_0}(3\varkappa - 1)\right)\varepsilon = 0 \qquad (3)$$

where

$$\omega_{1,2} = \sqrt{\left(3\varkappa P_\infty + \frac{2\sigma}{R_0}(3\varkappa - 1)\right)\bigg/\left(\rho\left(R_0^2 \pm \frac{R_0^3}{d_{12}}\right)\right)} \qquad (4)$$

and

$$\zeta_{1,2} = \frac{2\mu}{\rho\omega_0\left(R_0^2 \pm \frac{R_0^3}{d_{12}}\right)} \qquad (5)$$

where the sign ± is positive and negative for the in-and anti-phase modes, respectively, in the equation. Upon substation into

$$f = \frac{\omega_{1,2}}{2\pi}\sqrt{1 - \zeta_{1,2}^2} \qquad (6)$$

the resonant frequencies in the in- and anti-phase modes were calculated. We chose $R_0 = 6.3$ μm such that the bubble frequency as $d \to$ infinity coincided with the frequency of the single bubble shown in Fig. 1(b) ($f = 0.55$ MHz). The calculated oscillation frequencies at different $d$ in both modes are summarized in Fig. 3 via white lines. Although the calculations were performed using small-amplitude approximations, the calculation results reproduce our experimental data well. Thus, the extended Rayleigh-Plesset model is a valid framework for predicting the interactions between microbubbles vibrating even at large amplitudes. Moreover, the interaction between the two bubbles could be understood based on the pressure fluctuations produced by the oscillations of the bubbles. The vibrational frequency was highly susceptible to the bubble-bubble distance, with the slight distance tuning of less than 10 μm resulted in the modification of the vibration frequency from 0.5 MHz to more than 0.8 MHz, along with changes in the spatial profile of the oscillation. Therefore, by tuning the distance, the generated flow and sound waves around the bubble pair can be manipulated on demand, providing new dimensions for microfluidics to shape spatially and temporally regulated flow profiles.

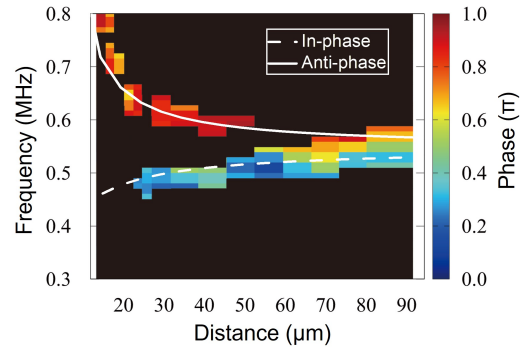

FIG.3. Color map showing the experimental results of oscillation frequencies and phase difference between the two bubbles at each distance between bubbles and each oscillation frequency. White lines show the theoretical calculation of the interaction between the two bubbles in both in-phase (dashed line) and anti-phase modes (solid line).

# Conclusion

In this study, we demonstrated the highly regulated vibrational coupling in laser-induced microbubble oscillations. Using an SLM to generate and control the distance between two laser beam spots and a 5 Mfps ultra-high-speed camera, we directly observed vibrational coupling between two high-speed vibrating microbubbles in real space and time. The coupled oscillation profiles were analyzed based on an extended Rayleigh-Plesset mode. Our findings revealed that the vibrational frequency of the microbubbles was highly sensitive to the inter-bubble distance, with slight adjustments in the distance causing significant changes in the oscillation frequency. Specifically, a variation of 10 μm in the distance between bubbles drastically changed the vibration frequency and oscillation state by more than 50%. This sensitivity underscored its potential for precise control in applications that require high-speed stirring, biosensing, and sonic communication in microfluidic environments.

We in the end emphasize the pervasive role of vibrational coupling across various disciplines in physical and chemical sciences. Vibrational coupling is crucial for understanding the molecular vibrations and acoustic/optical phonons in condensed matter physics. Although mechanical coupling through chemical bonds typically dominates such interactions, in certain instances, through-space dipole-dipole coupling results in the hybridization and delocalization of individual molecular vibrations [50-54]. The proposed microbubble system provides a structurally analogous platform to such through-space vibrational couplings, differing primarily in terms of the underlying mechanical or electrical mechanisms. Consequently, our ability to directly observe coupled vibrations in real space offers valuable insights into the dynamics of molecular vibrations and associated electromagnetic field flow. Furthermore, extensive knowledge developed in the field of molecular vibrations, including recent advancements in polaritonic coupling with cavities [55] and nano-cavities [56], are expected to inspire the development of fundamentally new approaches for controlling multi-bubble oscillations and the fluid dynamics surrounding them.

# Acknowledgments

This study was supported by the JST FOREST Program (Grant No. JPMJFR203N, Japan) and JSPS KAKENHI Grant No. 21H01784. It was also financially supported by a collaborative research project between Kyoto University and Mitsubishi Electric Corporation on Evolutionary Mechanical System Technology. We extend our heartfelt gratitude to Dr. Masaaki Sakakura for his invaluable guidance in refining the optical setup for simultaneous multipoint laser irradiation. We also extend our sincere appreciation to Dr. Takashi Kumagai (Institute for Molecular Science, Japan) for his insightful advice on analyzing and comprehending oscillator synchronization phenomena.

# Supplemental Material

The supplemental material contains Notes S1−S7: S1. Optical properties of the FeSi2 thin film;

S2. Experimental setup; S3. Typical microscopic images of the microbubble oscillation; S4. Comparison of the vibration frequency measured using the high-speed camera and that measured by the photomultiplier tube; S5. Relationship between the laser power and the maximum radius of vapor-rich bubbles; S6. Rayleigh-Plesset model for single microbubble oscillation; S7. Dependence of bubble oscillation frequency on distance between bubbles. The supplemental material also contains Movies S1−S4: S1. Single bubble oscillation; S2. Coupled bubble oscillation at d = 67.1 μm; S3. Coupled bubble oscillation at d = 27.6 μm; S4. Coupled bubble oscillation at d = 13.9 μm.